\documentclass[prb,amsmath,amssymb,twocolumn, superscriptaddress]{revtex4-1} 
\usepackage{url}
\usepackage{color}
\usepackage[english]{babel}

\usepackage{chemist}
\usepackage{graphicx}
\usepackage[version=3]{mhchem}

\begin{document}

\author{Ga\'etan Laurens}
\email{gaetan.laurens@univ-lyon1.fr}
\affiliation{Universit\'e Claude Bernard Lyon 1, CNRS, Institut Lumi\`ere Mati\`ere, F-69622, Villeurbanne, France}
\author{David Amans}
\email{david.amans@univ-lyon1.fr}
\affiliation{Universit\'e Claude Bernard Lyon 1, CNRS, Institut Lumi\`ere Mati\`ere, F-69622, Villeurbanne, France}
\author{Julien Lam}
\affiliation{Center for Nonlinear Phenomena and Complex Systems, Code Postal 231, Universit\'e Libre de Bruxelles, Boulevard du Triomphe, 1050 Brussels, Belgium.}
\author{Abdul-Rahman Allouche}
\affiliation{Universit\'e Claude Bernard Lyon 1, CNRS, Institut Lumi\`ere Mati\`ere, F-69622, Villeurbanne, France}

\title{Comparison of aluminum oxide empirical potentials from cluster to nanoparticle}

\begin{abstract}
Aluminum oxide nanoparticles are increasingly sought in numerous technological applications. 
However, as the nanoparticles grow during the synthesis, two phase transitions occur. At the nanoscale, numerical simulation of the stability of the alumina phases requires the use of empirical potentials that are reliable over a large range of system sizes going from a few atoms to several hundred thousand atoms. In this work, we confronted four different empirical potentials that are currently employed for bulk alumina. We found that only two of them are correct at the molecular level when compared to DFT calculations. Furthermore, the two potentials remain the best at the nanoscale as they reproduce one or two phase transitions that were observed experimentally: from amorphous solid to cubic crystal ($\gamma$) and from cubic to hexagonal ($\alpha$, \textit{i.e.} corundum) crystal.
\end{abstract}

\maketitle

\section{Introduction}

While the most stable crystal structure corresponds to the polymorph with the lowest bulk Gibbs free energy, at the nanoscale, surface Gibbs free energy becomes preponderant and can help to stabilize different polymorphs. As such, structural transitions towards metastable structures are observed when the particle size is decreased and a critical surface area is reached~\cite{Navrotsky2003}. Such crossover in polymorph stability was reported for several materials including 
\ce{Al2O3}~\cite{McHaleSci},
\ce{TiO2}~\cite{Zhang1998,Zhang2000,Zhu2005,Ranade2002},
\ce{ZrO2}~\cite{Garvie1965, Navrotsky2003,Chen2011},
\ce{Fe2O3}~\cite{Schimanke2000,Bomati2008},
\ce{Gd2O3}~\cite{Nicolas2006}
or \ce{GeTe}~\cite{Caldwell2010}.

As a bulk material, aluminum oxide is widely employed in the industry for its catalytic~\cite{brey1949surface,Ernst2004Jan}, mechanical,~\cite{Lukin2001Mar} and optical~\cite{Schembri2007Jun} properties.
At ambient conditions but also up to high temperature and pressure, alumina is usually found in the hexagonal phase ($\alpha-\ce{Al2O3}$), called corundum. 
At the nanoscale, amorphous solid~\cite{Tavakoli2013} and cubic crystal~\cite{Liu2010,Liu2011,Piriyawong2012,Lam2014,Ishizuka2018} were also synthesized and the stability of each phase depends mostly on the particle size. 
Ishizuka \textit{et al.} suggest that $\alpha-\ce{Al2O3}$ is not expected in condensation experiment because of the drastic difference between the crystal structure of the nucleus and the $\alpha$ phase~\cite{Ishizuka2016}. 
The phase transition from the crystal structure of the nucleus to the $\alpha$ phase during the growth would necessitate high temperature for such a refractory material.
In contrast, McHale \textit{et al.} have prepared $\alpha-\ce{Al2O3}$ and $\gamma-\ce{Al2O3}$ through topotactic decomposition of diaspore $\alpha$-AlOOH, and boehemite $\gamma$-AlOOH, respectively~\cite{McHaleJPCB}.  
Using calorimetry measurements, they addressed the polymorph stability and they showed that corundum is the most stable polymorph only for small surface areas below 125~m$^2/$g (above 12~nm in size)~\cite{McHaleSci}. 
Reducing the particle size and thus increasing the surface area, a first transition occurs from the hexagonal phase $\alpha$ to the cubic phase $\gamma$. Then, additional calorimetry measurements found another structural transition to an amorphous solid for a surface area larger than 370~m$^2/$g which corresponds to a critical size of 4~nm~\cite{Tavakoli2013}. From a numerical point of view, addressing such transition involves large scale simulations, which are inaccessible with DFT calculations. In this context, using empirical potentials requires their reliability and accuracy to hold from the molecular level to almost bulk-like systems. With oxide materials, the task exhibits additional complexities because of the oxygen bonding and the complex structural and stoichiometric landscapes.

First-principles calculations of aluminum oxide in the corundum phase were extensively employed to study bulk and surface properties~\cite{ruberto2003, manassidis1993, manassidis1994, marmier2004} as well as cluster structures~\cite{Sharipov2013, Li2012, Rahane2011, Lam2015, gu2015}. Interatomic potentials have also been used to study bulk and surfaces of aluminum oxide using structural optimization~\cite{mackrodt1987, mackrodt1989, hartman1989} and molecular dynamics simulations~\cite{salles2016, sun2006, adiga2006, blonski1993}. Numerous works have been achieved on the different polymorphs~\cite{lizarraga2011, krokidis2001, ruberto2003} including the cubic phase~\cite{blonski1993, Alvarez1994, Alvarez1995, Pinto2004}, the amorphous phase~\cite{adiga2006, Gutierrez2002, Gutierrez2011, Vashishta2008, hoang2004, nguyen2016}, and the liquid-like structure~\cite{ahuja1998, ansell1997, Vashishta2008}. All these studies investigated separately bulk, surfaces, nano-size particles, or clusters.
However, no calculations of alumina were performed over a large range of particle sizes. 
Recently, Erlebach \textit{et al.} combined first-principles methods (DFT) and molecular dynamics (MD) calculations to study the structural evolution of the hematite $\alpha-\ce{Fe2O3}$ particles~\cite{erlebach2015}, which present the same transition to the $\gamma$ crystal structure (maghemite) than alumina~\cite{Bomati2008}. 
An empirical potential was used to perform (i) a conformation research on $(\ce{Fe2O3})_n$ clusters with n$=1-10$, further refined by DFT optimizations and (ii) MD calculations on larger particles up to 5~nm.

Our work focuses on aluminum oxide for which we present a benchmarking of four empirical potentials that are already employed in the literature. These potentials have been selected because they were originally developed to reproduce the alumina phases. For each of them, geometry optimizations are performed on particles, from small clusters of a few atoms to nanoparticles larger than 12~nm, starting from both $\alpha$ and $\gamma$ bulk crystal structures. For the small clusters, results obtained for each empirical potential are compared to DFT-based calculations in order to assess their accuracy. Then, by probing the formation energy of the particles obtained, the crossovers in polymorph stability are deduced for each potential and compared with experimental data.

\section{Methods}

\subsection{Empirical potentials}

We selected four empirical potentials that were previously employed to study aluminum oxide and were developed respectively by  Alvarez \textit{et al.}~\cite{Alvarez1994,Alvarez1995},  Vashishta \textit{et al.}~\cite{Vashishta1990,Vashishta2008}, Woodley~\cite{Woodley2011}, and Streitz and Mintmire~\cite{Streitz1994}. Table~\ref{EQNPot} shows the mathematical formulation of each empirical potential with $r_{ij}$ being the distance between the atoms $i$ and $j$. The electrostatic contribution with attractive and repulsive terms is present in all of the potentials investigated. $q_{i}$ denotes the atomic charge of the atom $i$, while keeping in mind that the potentials are formatted using atomic units. 

\begin{table*}[th!]
\begin{tabular}{l|r}
Alvarez~\cite{Alvarez1994,Alvarez1995} & $V=\sum\limits_{i<j}\left[\frac{q_{i}q_{j}}{r_{ij}} + \frac{1}{p(\sigma_{i}+\sigma_{j})}\left(\frac{\sigma_{i}+\sigma_{j}}{r_{ij}}\right)^{p}\right]$ \hspace{0.5cm}(1)\\  \\ \hline\\ 
Vashishta~\cite{Vashishta1990,Vashishta2008} & $V=\sum\limits_{i<j}\left[ \frac{q_{i}q_{j}}{r_{ij}}e^{-r_{ij}/\lambda} + \frac{H_{ij}}{r_{ij}^{\eta_{ij}}} - \frac{D_{ij}}{r_{ij}^{4}}e^{-r_{ij}/\xi} 
- \frac{W_{ij}}{r_{ij}^{6}} \right] + \sum\limits_{i<j<k}\left[ R(r_{ij},r_{ik}) P(\theta_{jik}) \right]$ \hspace{0.5cm} (2)\\  \\ \hline\\ 
Woodley~\cite{Woodley2011} &  $V= \sum\limits_{i<j} \left[ \frac{q_{i}q_{j}}{r_{ij}} + \frac{A_{ij}}{r_{ij}^{12}}-B_{ij}\,exp\left(\frac{-r_{ij}}{\rho_{ij}}\right) -\frac{C_{ij}}{r_{ij}^{6}} \right]$\hspace{0.5cm} (3)\\  \\ \hline\\ 
Streitz~\cite{Streitz1994} &  $V= \left[ E_{0} + \sum\limits_{i}q_{i}\chi_{i} + \frac{1}{2}\sum\limits_{i,j}q_{i}q_{j}V_{ij} \right] + \left[\sum\limits_{i}F_{i}[\rho_{i}] + \sum\limits_{i<j}\phi_{ij}(r_{ij})\right] $ \hspace{0.5cm} (4)\\ 
\end{tabular}
\caption{Mathematical expressions of the empirical potentials employed.}
\label{EQNPot}
\end{table*}
\setcounter{equation}{4}

Firstly, Alvarez \textit{et al.} proposed the simplest model with a Coulomb term and a steric repulsion (Eqn. 1), where $\sigma_{i}$ is the ionic radius and $p$ the steric exponent.

Secondly, these terms were further developed in the Vashishta's model with the steric part expressed using the steric strengths $H_{ij}$ and the exponents $\eta_{ij}$. The Vashishta's potential includes also a charge-dipole term and a Van der Waals interaction contribution in the pairwise part (Eqn. 2), composed by their respective strengths $D_{ij}$ and $W_{ij}$. For the Coulomb and charge-dipole terms, an exponential decay was added to damp interactions at short distances. $\lambda$ and $\xi$ are their respective screening lengths. 
Moreover, a three-body part is added to constrain the stretching as well as the bending on the local crystal structure, with the functions $R(r_{ij},r_{ik})$ and $P(\theta_{jik})$, respectively.
Summations over $i$,$j$ and $k$ are limited to the local environment of each atom, using a cutoff $r_{0}$ on the distances. Another cutoff $\theta_{0}$ on the angles contributes to favor the $\alpha$ crystal structure.

Thirdly, the so-called Woodley's potential, which was also developed by Guti\'errez \textit{et al.} for a prolific work on the liquid~\cite{Gutierrez2000} and amorphous~\cite{Gutierrez2002,Gutierrez2011} phases of alumina, consists on the Coulomb contribution and the Buckingham potential terms, including interaction parameters B$_{ij}$, C$_{ij}$, and $\rho_{ij}$. However, some divergence issues at short distances led Woodley to add a Lennard-Jones repulsive term in $r^{-12}$ (Eqn. 3). 

At last, Streitz and Mintmire developed a potential including a variable charge electrostatic potential (ES) in addition to an empirical potential embedded-atom method (EAM). The first contribution of equation 4 describes the atomic energies and the electrostatic interactions. $E_{0}$ includes the neutral atomic energy and nuclear terms, which are independent of the atomic charges. $\chi_{i}$ is composed by the atomic electronegativity, and nuclear-electronic interaction integrals. $V_{ij}$ contains the atomic hardness and the electronic interaction term. 
The second contribution is the EAM potential. $F_{i}[\rho_{i}]$ is the energy required to embed an atom in the local electron density $\rho_{i}$, which is build as a linear superposition of atomic density functions. $\phi_{ij}(r_{ij})$ is the pair potential required for the short-range repulsive description of the pair interactions. The main advantage of this approach is to take into account the modification of the local atomic charge due to the environment of each atom. The electrostatic compound includes the variable atomic charges while the EAM term takes into account the local environment such as the under-coordinated atoms on the surface. 

Following the functional forms of these potentials, their original developers carried out two different parameterizations. On the one hand, for the Alvarez's potential, ab initio calculations on \ce{Al(OH)x} with x=2-6 were performed and the parameters were adjusted in order to reproduce the most common polyhedra of the alumina phases, \textit{i.e.} tetrahedral and octahedral coordination. On the other hand, the other potentials were parameterized to reproduce the bulk structural properties of $\alpha$-\ce{Al2O3} (lattice constants, cohesive energy, bulk modulus, and elastic constants).

The empirical potentials have been implemented in a home-made code in order to perform all the calculations shown in this study.

\subsection{Molecular cluster \ce{(Al2O3)_n} with $n\leq8$}

Conformation research was performed for molecular clusters $(\ce{Al2O3})_n$ ($n=1-4, 6, 8$) following a 4-step process:
\begin{enumerate}
\item For the small clusters $(\ce{Al2O3})_n$ ($n=1-3$), Density Functional Theory (DFT) calculations were performed by using the Gaussian software.
2000 random geometries were generated and optimized by using successively a 6-31G* and a 6-311+G* basis sets with the B3LYP functional.
Between each step, the similar geometries were removed. 
A first set of the most stable conformers according to DFT were thus found.

\item In parallel, conformation research was also carried out by starting with random configurations that were then optimized with the empirical potentials. 
Calculations were performed for \ce{(Al2O3)_n} using $n \times 10^4$ random geometries for $n={1, 2, 3, 4, 6}$, and $6 \times 10^4$ for $n=8$.
Indeed, the optimization using potentials is drastically less time-consuming than DFT, which enables to work with larger sizes ($n$) and a large number of random geometries.
However, only the Alvarez's and Streitz's potentials were used for $n={6, 8}$ because they appeared to be the most reliable.
After removing the similar structures, the remaining geometries were ordered by their energy. See Supplemental Material at [URL will be inserted by publisher] for Figures S1-S6 displaying the five most stable clusters for each potential.

\item In addition, the lowest-energy clusters found at step 2 with the empirical potentials were further refined by DFT using a B3LYP/6-311+G* functional/basis set combination.
In particular, the ten and the thirty lowest-energy structures were selected for $n={1, 2, 3}$ and for $n={4, 6, 8}$, respectively. 
This allowed us to find additional structures that we have not obtained when using only DFT (See Supplemental Material at [URL will be inserted by publisher] for Figures S1-S6). 

\item Finally, the five isomers obtained from DFT with the lowest energies (step 3) were recalculated using each potential. Following this step, it appeared that isomers with the lowest energies already found at step 2 were retrieved for each potential. For a given potential, the isomers that did not appear with a certain potential were unstable or located at a much higher energy levels. We found missing structures only for the Streitz's potential for cluster sizes larger than $n = 3$ (Flagged with a double asterisk in Figures \ref{fig2} and S5-S6).
\end{enumerate}

The obtained DFT geometries are comparable to previous studies~\cite{Lam2015,Sharipov2013,Li2012,Rahane2011}.
In particular, isomers reported by Li~\textit{et al.}~\cite{Li2012} and their relative energies are perfectly consistent with the ones reported in this work. Only a few studies addressing the largest clusters provided results that to do not exactly match with our DFT results on (\ce{Al2O3})$_{8}$ clusters~\cite{Rahane2011, Gobrecht2018}.

\subsection{Nanoparticles $(\ce{Al2O3})_n$ nanoparticles from n~=~50 to n~=~20~000}

$(\ce{Al2O3})_n$ nanoparticles from n~=~50 to n~=~20~000 were optimized with the empirical potentials for three different crystal structures (See Supplemental Material at [URL will be inserted by publisher] for Fig. S7 where their respective diameters are displayed). 
For each crystal structure, the primitive lattice cell was duplicated to generate large supercells of $\alpha$ and $\gamma$ phases. $\alpha$-\ce{Al2O3} supercell was built from the hexagonal crystal structure of corundum (52648 ICSD file~\cite{Lewis1982}), characterized by the lattice parameters a = b = 4.76~\AA~and c = 12.99~\AA, and the space group $R\overline{3}c$.
$\gamma$-\ce{Al2O3} is described by a cubic crystal structure (04-005-4662 ICDD file~\cite{Guse1990}) with a lattice parameter a = 7.948~\AA~and with the space group $Fd\overline{3}m$. 
The $\gamma$ phase presents a defect spinel structure for which the Site Occupancy Factor (SOF) cannot reach unity for one or more aluminum sites to fulfill the stoichiometry of alumina \ce{Al2O3}.
There are several available crystallographic information files and the position of the vacancies at the octahedral or tetrahedral sites is still under debate in the scientific community.
To create the $\gamma$ nanoparticles, at first an ideal \ce{Al3O4} spinel structure is built. 
Then, aluminum atoms are removed to get the right stoichiometry. 
Pinto \textit{et al.}~\cite{Pinto2004} studied all possible configurations for the position of the vacancies and concluded that the octahedral site led to the most stable primitive cell.
Therefore, we prepared two configurations: (1) $\gamma_{Oh}$ with only aluminum vacancies on octahedral sites and (2) $\gamma_{TdOh}$ where aluminum atoms are randomly removed. 
Finally, supercells of hundreds of thousands of atoms were built, then particles of the desired size could be cut directly from them. 
The three crystal structures are named $\alpha-(\ce{Al2O3})$, $\gamma_{Oh}-(\ce{Al2O3})$ and $\gamma_{TdOh}-(\ce{Al2O3})$. 
It is important to mention that these notations correspond to the starting bulk crystal.

\subsection{Molecular dynamics calculations} 
Molecular dynamics (MD) calculations have been also performed with the Alvarez's potential in order to confirm results that were obtained by optimization. From unrelaxed crystal structures, the MD calculations are run in three stages with a step time of 0.5 fs: (i) The temperature of the system is increased from 0 to 300 K during 1 ps. (ii) The system is equilibrated at 300 K during 1ps. (iii) A production stage runs during 10 ps. The Bussi thermostat is employed to fix the temperature\cite{bussi2007}.

\subsection{Analysis tools}

\begin{figure}[h]
\centering
  \includegraphics[width=8.6cm]{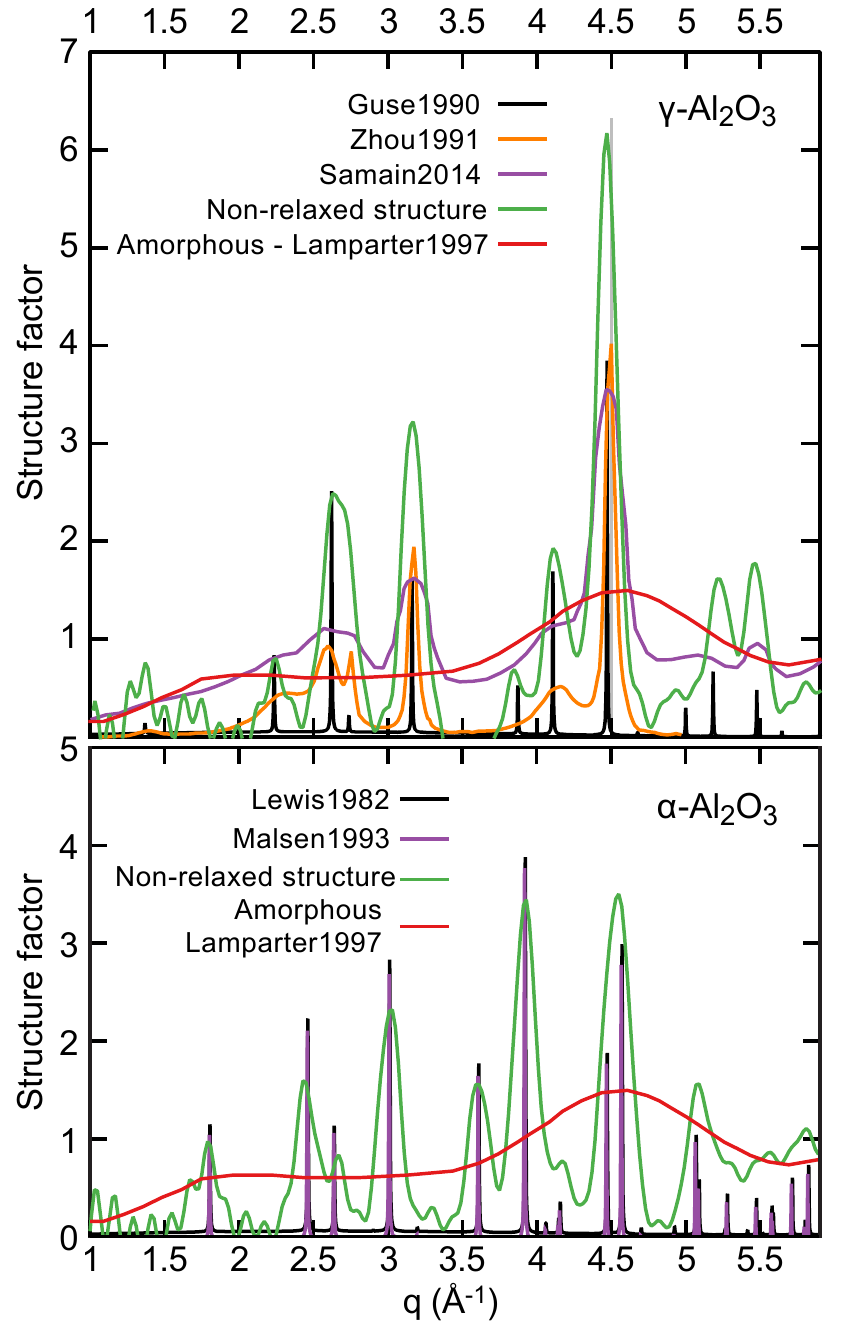}
  \caption{Structure factors S of the as-build supercells (green curves) compared with experimental reference data for $\gamma$~\cite{Guse1990, Zhou1991, Samain2014} (top) and $\alpha$~\cite{Lewis1982, Maslen1993} (down) phases, as well as for the amorphous phase~\cite{Lamparter1997}.}
  \label{fig1}
\end{figure}

The structure of the resulting nanoparticles is characterized with the Structure Factor (S) and the Coordination numbers ($n_c$). These two quantities are related to the Radial Distribution Function (RDF)~\cite{Gutierrez2000, Gutierrez2002, Vashishta2008}. These quantities reliably quantified the crystallinity and the structural organization of the particle at long and short distances. 

\textbf{Radial Distribution Function $g(r)$.}
The Radial Distribution Function describes how the density varies as a function of distance from a reference atom.
The RDF gives insight about the long-distance order which can be found in a crystalline structure.
The quantities below are defined for a binary system with atoms of species $a$ and $b$ ($a$, $b$ = Al, O here).
The partial pair-distribution function $g_{ab}(r)$ (Eqn \ref{eq:gab}) is calculated from the average number of atoms $b$ surrounding an atom $a$ in a shell of thickness $\Delta r$ (between $r$ and $r+\Delta r$)~\cite{Vashishta2008}.
\begin{equation}\label{eq:gab}
g_{ab}(r) = \frac{<n_{a,b}(r,r+\Delta r)>}{4\pi r^{2}\Delta r \rho c_{b}}
\end{equation}
where $\rho$ is the material density and $c_{b}$ is the concentration of the atom $b$.
The brackets denote an ensemble average over all $a$ atoms.
The concentration $c_{b}$ of the atom $b$ (respectively $c_{a}$) is the ratio between $N_b$ the number of atoms $b$  (respectively $N_a$) and the total number of atoms $N=N_a+N_b$.

Summations over partial functions $g_{ab}(r)$ lead to the total radial pair distribution g(r), as well as to the neutron  distribution function $g_{n}(r)$ given by:
\begin{equation}\label{eq:g}
g(r) = \sum_{a,b} c_{a}c_{b}g_{ab}(r) 
\end{equation}

\begin{equation}\label{eq:gn}
g_{n}(r) = \frac{\sum_{a,b} c_{a}b_{a}c_{b}b_{b}g_{ab}(r)}{(\sum_{a} c_{a}b_{a})^{2}}
\end{equation}
with $b_{a(b)}$ the coherent neutron scattering cross section of the atom a(b). \\

\textbf{Coordination Number $n_c$.}
The number of the first-neighbours $b$ surrounding an atom $a$ corresponds to the coordination number for this atom.
It can be calculated from the partial RDF :
\begin{equation}\label{eq:n}
n_{ab}(r) = 4\pi \rho c_{b} \int_{0}^{R_{max}} r^{2}g_{ab}(r)dr
\end{equation}
The cutoff radius $R_{max}$ was chosen to be fixed for each type of bonds in order to compare the potentials: R$_{max}$(Al-Al) = 3.10~\AA~$>$~R$_{max}$(Al-O) = 2.43~\AA~$>$~R$_{max}$(O-O) = 1.76~\AA.
This quantity gives an insight on the coordination geometry which characterizes the crystal structure at a short distance
if one focuses on the cations ($a$=Al and $b$=O).\\

\textbf{Structure Factor $S(q)$.}
The static structure factor $S$ allows us to observe the long-distance order of a crystal.
$S(q)$ is the main parameter that defines the intensity of the peaks in the x-rays or neutron diffractograms.
Consequently, the structure factor helps to compare the theoretical calculations to the experimental results.
$S(q)$ is obtained from the Fourier transform of the partial RDF:
\begin{equation}\label{eq:Sab}
\begin{split}
S_{ab}(q) = \delta_{ab} + 4\pi \rho (c_{a}c_{b})^{1/2} \times \\
\int_{0}^{R}[g_{ab}(r)-1]\frac{r^{2}sin(qr)}{qr}\frac{sin(\pi r/R)}{\pi r/R}dr
\end{split}
\end{equation}
where $R$ is the cutoff distance chosen here as half the simulation box length.
The window function $\frac{sin(\pi r/R)}{\pi r/R}$ has been introduced to reduce the termination effects~\cite{Gutierrez2002}.
From this partial static structure factor, we can calculate the x-rays and neutron structure factors:
\begin{equation}\label{eq:Sx}
S_{X}(q) = \frac{\sum_{a,b} (c_{a}c_{b})^{1/2}f_{a}f_{b}S_{ab}(q)}
{\sum_{a} c_{a}f_{a}^{2}}
\end{equation}
\begin{equation}\label{eq:Sn}
S_{n}(q) = \frac{\sum_{a,b} (c_{a}c_{b})^{1/2}b_{a}b_{b}S_{ab}(q)}
{\sum_{a} c_{a}b_{a}^{2}}
\end{equation}
with $f_{a(b)}$ the x-rays form factors~\cite{Gutierrez2002}. \\

The reliability of our supercells and the implementation of the analytic tools were cross-checked with different references by analyzing the static structure factor S (see figure~\ref{fig1}). Maslen \textit{et al.}~\cite{Maslen1993} worked on the $\alpha$ phase while the groups of Zhou~\cite{Zhou1991} and Samain~\cite{Samain2014} studied the $\gamma$ phase. As shown in figure~\ref{fig1}, green curves of our structures fit perfectly with the literature data. A red curve from the amorphous phase measured by the group of Lamparter~\cite{Lamparter1997} is added for comparison.

\section{Results}

\subsection{Comparison between empirical potentials and DFT calculations}

\begin{figure*}[!ht]
\centering
  \includegraphics[width=18cm]{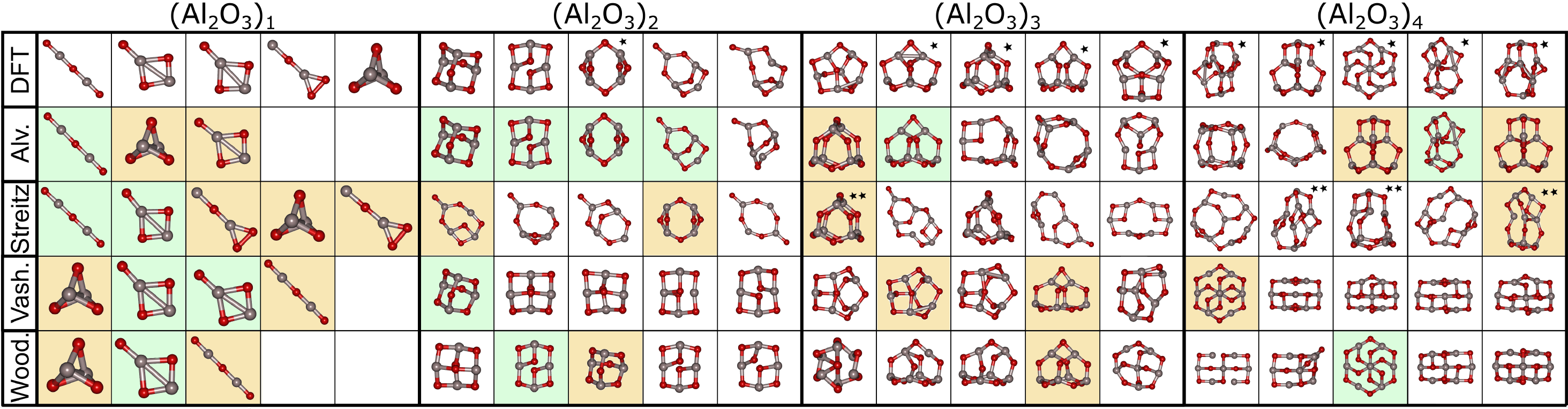}
  \caption{The clusters $(\ce{Al2O3})_{n}$ with $n=1-4$ calculated using DFT are compared to those calculated with the potentials. The single asterisk represents the DFT structures found after step 3, while the double asterisks shows the isomers found after step 4. Green and orange squares represent a cluster found using both DFT and potentials. The green color indicates the same rank between the two methods. Aluminum and oxygen ions are in grey and red, respectively.}
  \label{fig2}
\end{figure*}

The conformation research using the potentials results in diverse structures, displayed in Figure \ref{fig2} for $(\ce{Al2O3})_{n}$ with $n=1-4$. 
While being the most complex potentials, the Streitz's, Vashishta's and Woodley's have difficulties to predict the lowest-energy isomers with structures close to those obtained by DFT.
In particular, the Vashishta's potential favors compact 3D-structures with regular bond lengths and angles constrained by the stretching and bending terms in the three-body part of the potential. 
For the Woodley's potential, the very compact structures are due to the overestimated partial charges. 
By contrast, the Streitz's potential favors planar structures for the smallest structures. This difficulty at small sizes has been explained elsewhere~\cite{zhou2004}.   
Surprisingly, while being the most simple potential (See Eqn. 1), the Alvarez's potential is also quite efficient at predicting the lowest-energy geometries for $n=1-4$.

By constrast, the complexity of the DFT structures is not retrieved for the largest clusters by using the Alvarez's potential, whereas the structures found with the Streitz's potential appear closer (See Supplemental Material at [URL will be inserted by publisher] for Figures S5 and S6 for $(\ce{Al2O3})_{n}$ with $n=6,8$, respectively). Indeed, by increasing the size of the clusters, isomers found with this potential appear closer to the DFT ones, even though they are slightly deformed.

Interestingly, the use of such potentials allows us to recover missing clusters not found using the DFT, as shown by the clusters with a star displayed in Figure \ref{fig2}.
Overall, such results show that empirical potentials can be successfully employed as a starting optimization tool thus limiting the use of computationally expensive DFT calculations~\cite{Woodley2011, erlebach2015, Zhang2015}.

\subsection{Phase transition in nanoparticles}

\begin{figure*}[!ht]
\centering
  \includegraphics[width=16cm]{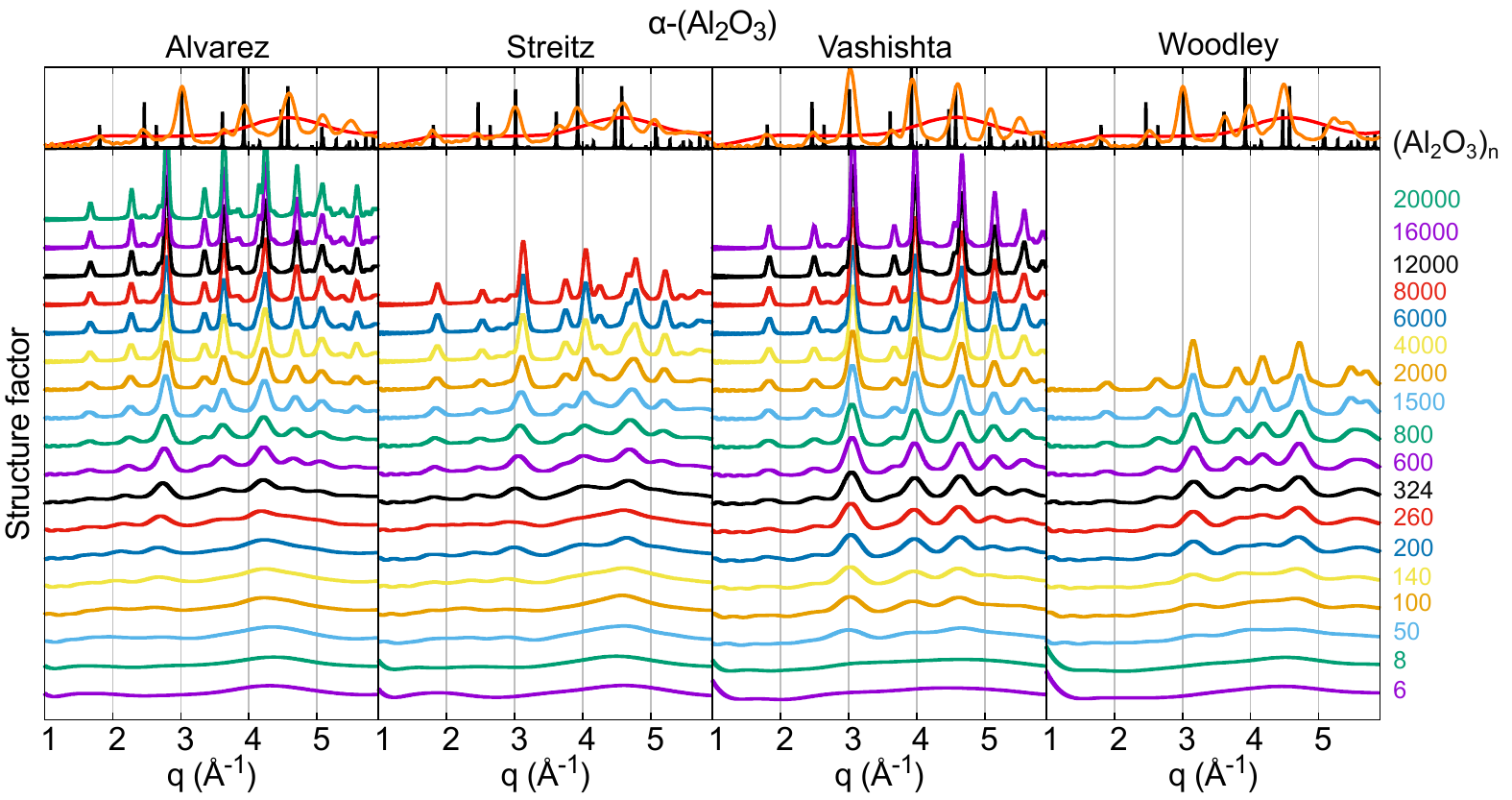}
  \caption{(a) Structure factors S computed for the relaxed $\alpha-(\ce{Al2O3})_{n}$ nanoparticles using all potentials. 
  For each curve, the $n$ value is displayed on the right. 
  Experimental data of $\alpha-(\ce{Al2O3})$ bulk~\cite{Lewis1982}(black curve) and $a-(\ce{Al2O3})$ bulk~\cite{Lamparter1997} (red curve) are displayed on top plots. The orange curve shows the homothety corrections of S curves designed for each potential on the $(\ce{Al2O3})_{1500}$ nanoparticle. 
  Similar plots presenting the particles evolution from the three crystal structures for each studied potential are available. See Supplemental Material at [URL will be inserted by publisher] for Fig. S8. }
  \label{fig3}
\end{figure*}

\begin{figure}[!ht]
\centering
  \includegraphics[width=8.6cm]{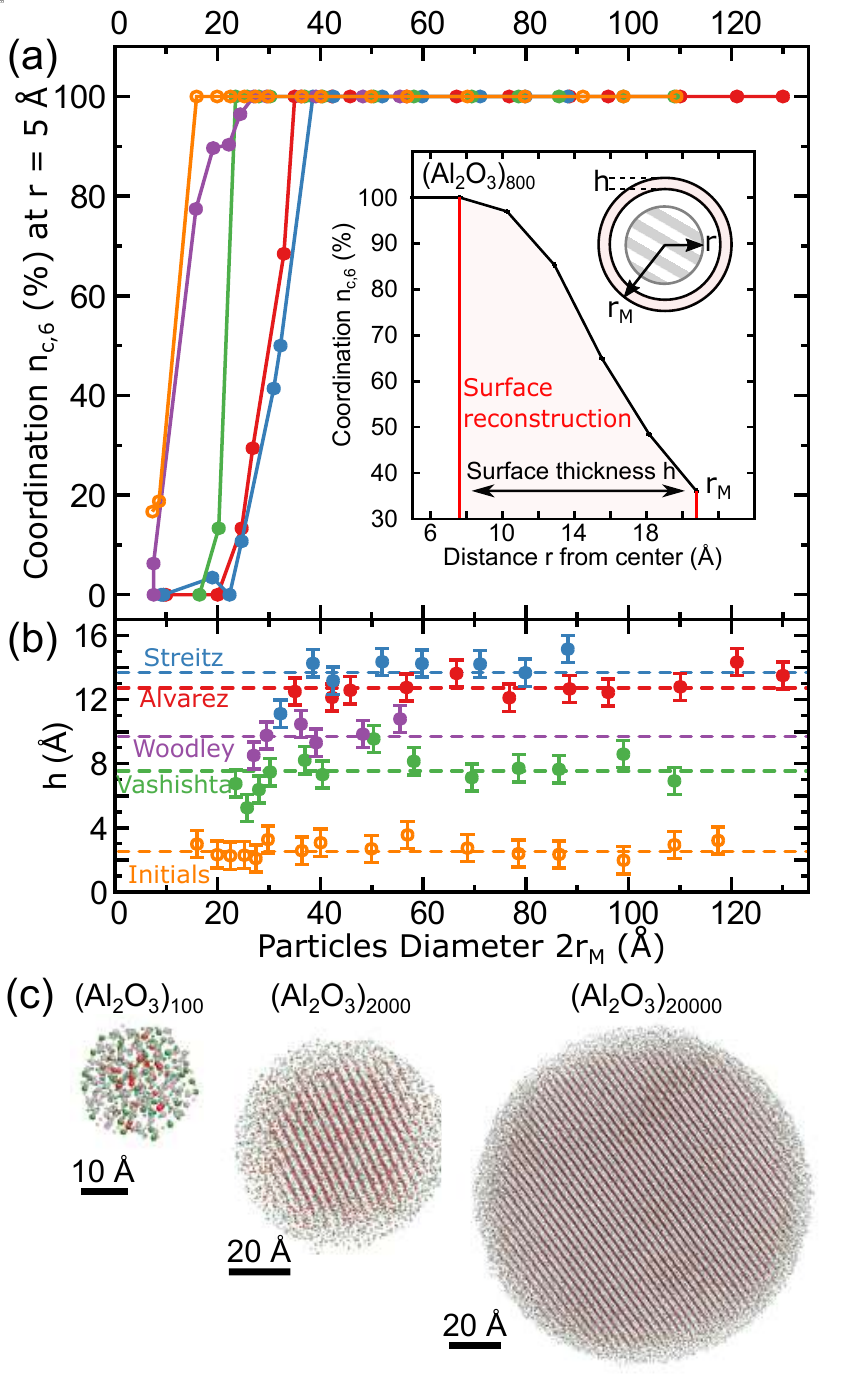}
  \caption{(a) Evolution of the percentage of the six-coordinate aluminum
cations ($n_{c,6}$) with the increasing particles radius $r_M$ calculated in the particles core (in a sphere of 5~\AA$\,$ radius). (Inset) Particles are subject to a surface reconstruction calculated from the $n_{c,6}$ distribution profile considering the distance $r$ from the center of the particle. (b) The thickness of the surface shell $h$ appears to be independent of the particle's size. (c) Final $\alpha$-\ce{Al2O3} nanoparticles optimized by the Alvarez's potential, revealing the growing crystal structure core surrounded by a constant amorphous shell.} 
  \label{fig4}
\end{figure}

After assessing the accuracy of the empirical potentials for the small clusters, we continued by studying larger systems with a particular focus on the structural transitions that were observed experimentally, \textit{i.e.} amorphous solid $\rightarrow$ cubic crystal ($\gamma$) $\rightarrow$ hexagonal crystal ($\alpha$, \textit{i.e.} corundum. \\

\textbf{Phase transition at global order.}\\
Figure \ref{fig3} compares the structure factors S of the particles obtained starting from $\alpha-(\ce{Al2O3})_n$ [See Supplemental Material at [URL will be inserted by publisher] for $\gamma_{Oh}-$ and $\gamma_{TdOh}-\ce{(Al2O3)}$ results in Fig. S8]. Experimental data from bulk corundum~\cite{Lewis1982} (black curve) and amorphous phase~\cite{Lamparter1997} (red curve) are displayed as a reference in the upper plots.
Regardless of the potential, the static structure factor shows a transition from flat bands at small sizes to sharp peaks at the larger ones. The smallest particles present a large band centered around 4.5~\AA$^{-1}$ as seen in the experimental curve of the amorphous phase~\cite{Lamparter1997}. By increasing the size, one can observe the emergence of predominant peaks that become sharper thus characterizing the crystal organization within the system. 
After correcting the small shift caused by homothety in the bond lengths (orange curve in the upper plots of figure~\ref{fig3}), a very good agreement is observed between the peaks positions and experimental results of the $\alpha$ bulk phase~\cite{Lewis1982}. This evolution which appears with the first peaks in height and width indicates a progressive phase transition from an amorphous phase to a crystal phase. The amorphous to crystal transition is found to start for $(\ce{Al2O3})_{100}$ particle (about 20~\AA) using the Vashishta's potential, and for  $(\ce{Al2O3})_{200}$ particle (about 24~\AA) using the Woodley's potential. Similarly to the smaller clusters, the three-body term of the Vashishta's potential favors crystal structure by constraining bond lengths and angles and consequently leads to a transition at smaller particle size than in experiments. Using the Alvarez's and the Streitz's potentials, the transition occurs from the $(\ce{Al2O3})_{324}$ particle, \textit{i.e.} respectively at around 35~\AA$\,$ and 32~\AA$\,$ which is in a much better agreement with the experimental value found at 40~\AA~ through calorimetry measurements~\cite{Tavakoli2013}. Until now, we have only focused on results obtained when starting with the $\alpha$ crystal configuration. However, it appears that the amorphous to crystal transition is found at approximately the same size for each potential regardless of the starting crystal structure [See Supplemental Material at [URL will be inserted by publisher] for Fig. S8].\\

\textbf{Phase transition at local order.}\\
For a more quantitative picture, the phase transition may be identified at local order using the coordination numbers n$_c$ of the Al atoms. Figure~\ref{fig4}(a) shows the evolution of the percentage of the six-coordinate aluminum cations n$_{c,6}$ (octahedral sites coordination) in the core of each particle when starting from $
\alpha-(\ce{Al2O3})$. A transition is observed for a size range which corresponds to a sharp increase of n$_{c,6}$. 
As such, the transition from amorphous (0~\%) to crystal (100~\%) is obtained at positions similar to what was found using the structure factor. 
While the phase transition toward the $\alpha$ phase occurs at lower size with the Vashishta's and the Woodley's potentials, \textit{i.e.} between 16 and 23 \AA~for the former and between 7 and 27~\AA~for the latter, the crystallization of the nanoparticles seems to range from around 20~\AA$\,$ to 35~\AA$\,$ and 38~\AA$\,$ with the Alvarez's and Streitz's potentials, respectively.
Ultimately, we thus confirm that Vashishta's and the Woodley's potentials lead to a transition for too small nanoparticles, while the results obtained with the Alvarez's and Streitz's potentials fit well with the experimental results.

Then, we employed the coordination distribution along the nanoparticle structure to understand further the observed amorphous to crystal transition. Inset of Figure \ref{fig4}(a) shows the distribution profile of n$_{c,6}$ as a function of the particle distance to its center in the case of the $(\ce{Al2O3})_{800}$ particle calculated with the Streitz's potential. Only at the core of the particle, all aluminum ions have the bulk coordination number thus showing that the particle surface is made of an amorphous shell. Its thickness, denoted $h$, is defined as the value for which we no longer have a 100 \% of aluminum ions in the bulk symmetry. From figure~\ref{fig4}(b), the surface thickness seems independent of the particle size but varies slightly from around 7.5~\AA~ to 13.8~\AA~ according to the used potential. This observation reveals that structures with radius $r_{M}$ smaller than this surface thickness $h$ are deformed as a whole and appear in the amorphous phase, as can be seen with the snapshots of three particles displayed in figure~\ref{fig4}(c). 
Ultimately, the amorphous to crystal transition occurs when the shell thickness exceeds the particle radius. \\

\begin{figure}[h]
\centering
  \includegraphics[width=8.6cm]{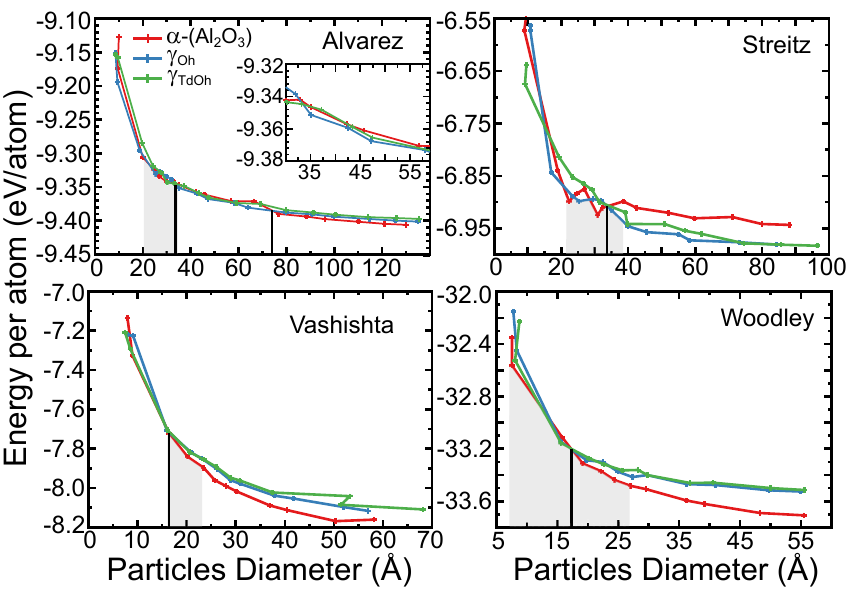}
  \caption{
  Energies per atom as a function of the nanoparticles diameters for each potential. Red, blue and green curves correspond, respectively, to the particles starting from the $\alpha$, $\gamma_{Oh}$, and $\gamma_{TdOh}$ crystal structures. Vertical black lines show the sizes of the phase transitions estimated from the S curves (see Figures~\ref{fig3} and S8). }
  \label{fig5}
\end{figure}

\textbf{Polymorph stability.}\\
Finally, we sought to clarify the stability regions of each crystal structure in order to evaluate the phase transitions predicted by every potential. 
In Figure \ref{fig5}, energy per atom is plotted as a function of the particle size for each starting phase and for each potential. Phase transition ranges estimated in figure~\ref{fig4}(a) are reported within the grey areas. Additionally, crossovers of the energy curves, indicated by the vertical black lines, give refined values of the crossover sizes which are compared to the calorimetry values measured by McHale \textit{et al.}~\cite{McHaleSci} and Tavakoli \textit{et al.}~\cite{Tavakoli2013}.

The $\alpha$ phase appears as the most stable phase for the Vashishta's and Woodley's models for every size, which again demonstrates that these two potentials cannot be employed for nanoscale simulations.
By contrast, for intermediate sizes, the Alvarez's and Streitz's potentials both predict that the $\gamma$ phase is energetically favorable as in experiments. Interestingly, these two potentials favor the $\gamma_{Oh}$ configuration instead of its $\gamma_{TdOh}$ counterpart, which is the expected structure according to Pinto \textit{et al.}~\cite{Pinto2004}. 

Regarding the second phase transition ($\gamma \rightarrow \alpha$), we cannot argue on the ability of the Streitz's potential to predict the transition at larger sizes, due to the computational cost. 
By contrast, the Alvarez's potential is able to predict, at least qualitatively, the stability of the polymorphs on the full particle size range. The transition is predicted at a smaller size with respect to the experimental value~\cite{McHaleSci}, 77~\AA$\,$ instead of  117~\AA$\,$.

\begin{figure}[!ht]
\centering
  \includegraphics[width=6cm]{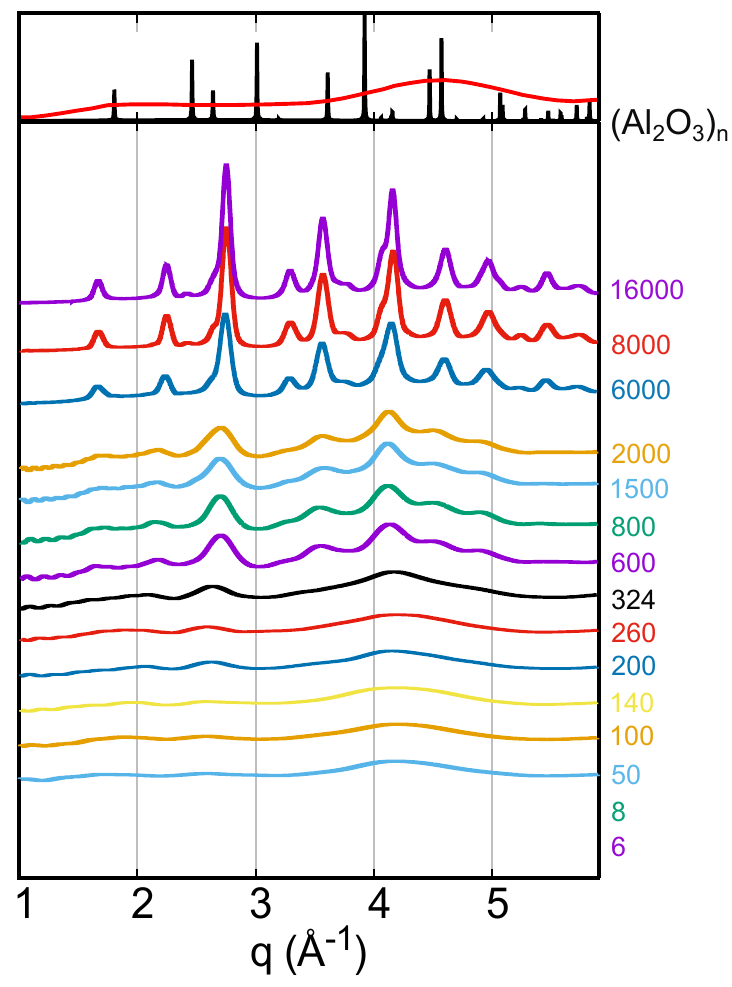}
  \caption{ Structure factors S of the $\alpha-(\ce{Al2O3})_{n}$ nanoparticles computed by MD using the Alvarez's potential. For each curve, the $n$ value is displayed on the right. Experimental data of $\alpha-(\ce{Al2O3})$ bulk~\cite{Lewis1982}(black curve) and $a-(\ce{Al2O3})$ bulk~\cite{Lamparter1997} (red curve) are displayed on top. } 
  \label{fig6}
\end{figure}

\begin{figure}[!ht]
\centering
  \includegraphics[width=8.6cm]{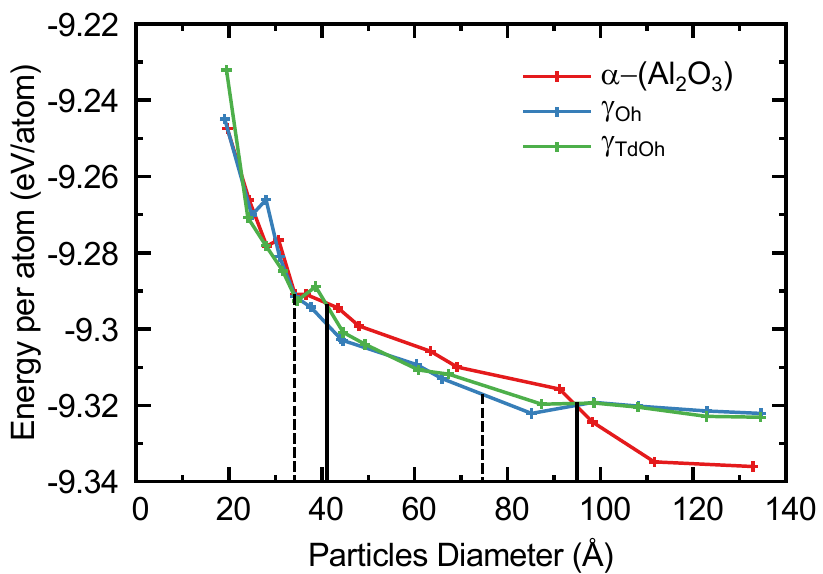}
  \caption{Energies per atom as a function of the nanoparticles diameters are deduced from the MD calculations using the Alvarez's potential. Red, blue and green curves correspond, respectively, to particles starting from the $\alpha$, $\gamma_{Oh}$, and $\gamma_{TdOh}$ crystal structures. The estimated sizes of the phase transitions are represented for the MD calculations (solid black lines) and the optimization (dashed black lines).} 
  \label{fig7}
\end{figure}

\textbf{Molecular dynamics calculations.}\\
The results of the optimization calculations using the Alvarez's potential appear consistent with the experimental data. In order to check the validity of these results, we performed molecular dynamics (MD) calculations using the Alvarez's potential. Figure \ref{fig6} displays the structure factors of the $\alpha$ particles obtained by MD. The structures of the particles computed by MD do not appear modified compared to the optimized particles (see Fig. \ref{fig3}). However, the phase transition from the amorphous phase to the $\alpha$ phase occurs at larger size, \textit{i.e.} at n=600 (around 43~\AA) with MD instead of n=324 (around 35~\AA) with optimization. \\
To go further, we displayed the energy per atom as a function of the particle sizes (see Fig. \ref{fig7}) and found a similar evolution. Indeed, two phase transitions are visible at 41~\AA~and at 95~\AA, corresponding to the amorphous $\rightarrow$ $\gamma$ transition and the $\gamma$ $\rightarrow$ $\alpha$ transition. These new values are appreciably closer to the experimental values~\cite{McHaleSci,Tavakoli2013} of 41~\AA~and 117~\AA~than those obtained by optimization. \\

\section{Conclusion}

\begin{figure}[!ht]
\centering
  \includegraphics[width=8.6cm]{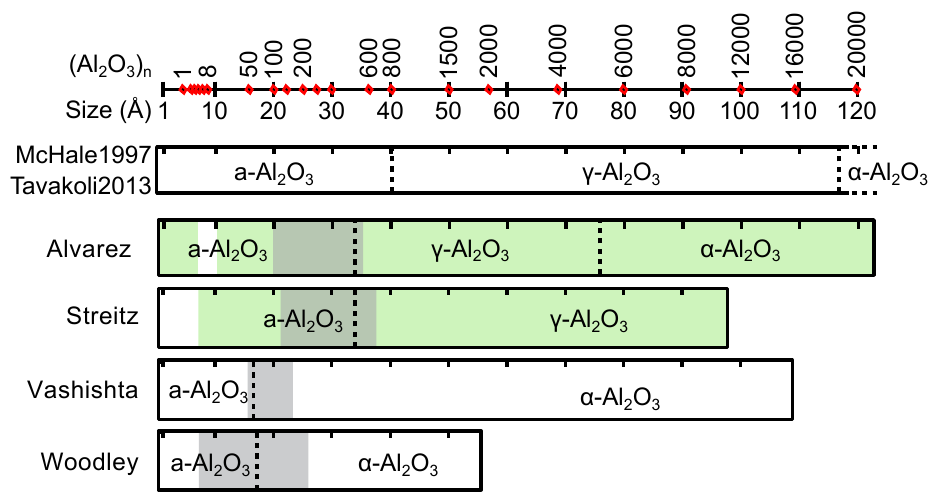}
  \caption{Overview of regions corresponding to the successive stable polymorphs calculated for each potential over the large range of particles' sizes. Crystallisation ranges estimated in figure~\ref{fig4} are reported within the grey areas. The dashed black lines correspond to the crossovers in polymorph stability deduced from the energies per atom displayed in figure~\ref{fig5}. They are compared to those deduced from calorimetry measurements in the top frame~\cite{McHaleSci,Tavakoli2013}. The computed particles' sizes are displayed along the scale bar, also graduated by their units numbers $n$. Please note that due to computational time limitation, the larger n values (12000, 16000, 20000) have converged only for the Alvarez's potential. Green areas show the size range where each potential describes accurately the particles according to their structures and their relative phase stability. } 
  \label{fig8}
\end{figure}

To summarize, we compared four empirical potentials in the framework of particle sizes ranging from a few atoms to 12 nm. The results obtained with each potential are summarized in Figure \ref{fig8} where the potential accuracy is represented by a green area along with its respective size range. In particular, three main features were investigated. \\
Firstly, we showed that, when compared to DFT results, the most complex potentials, naming Streitz's, Vashishta's and Woodley's potentials, have difficulties to recover DFT results on the smallest alumina molecules (\ce{Al2O3})$_n$ with $n=1-4$. For this size range, the Alvarez's potential shows the best agreement with DFT results. However, the Streitz's potential becomes more efficient as the clusters size increases (for $n={6, 8}$). Empirical potentials can be employed in a conformation search as a first optimization step before DFT calculations~\cite{erlebach2015,Gutierrez2011,Escatllar2019}.

Secondly, the amorphous to crystal transition occurring at intermediate sizes was found for all the potentials. We then demonstrated that such transition does not depend on the starting crystal structure and simply results from the lack of local order at nanoparticle surface shell induced by the atomic reconstruction of the bulk-to-surface cut.  

Thirdly, we compared the energy for each crystal structure and showed that only the Alvarez's and the Streitz's potentials managed to find the correct polymorph at the intermediate sizes. Surprisingly, the Alvarez's can also predict the solid-solid transition from $\gamma$ to $\alpha$ structures. Even though the full range of sizes was not explored using the Streitz's potential, it appears to be particularly promising for the large nanoparticles.
Finally, by performing MD calculations with the Alvarez's potential, we confirmed the results obtained by optimization but we also found a better agreement with the experimental findings. Such qualitative agreement again shows the efficiency of the Alvarez's potential while being the simplest in its formulation.

Altogether, this work shows that it is difficult to construct an empirical potential that remains accurate from the molecular level to nanometric scale. While it is commonly assumed that adding complexity in the functional form of the potential enables to improve its accuracy, the aluminum oxide example indicates that a much simpler potential may sometimes be able to compete with these complex potentials, \textit{e.g.} the Streitz's potential. Indeed, Alvarez's potential is the only one that was parameterized not on the bulk phase but on clusters obtained through DFT calculations. 
On the whole, this works shows an example where improvement of the potential can be made not by increasing its functional form but simply by using a more appropriate training database. Finally, this work is a first step towards the understanding of the phase transition occurring during the synthesis of nano-\ce{Al2O3}. Indeed, by comparing those different empirical potentials, it appears that the Alvarez's potential is the best candidate for future studies that include free energy barrier calculations~\cite{Calvo2011Jul,Pavan2015Nov,Bechelli2017Sep} and simulations of inert gas quenching~\cite{Singh2014Jun,Zhao2015Jan,Akbarzadeh2017Aug,Forster2019Oct}.

\section{Acknowledgement}

JL acknowledges financial support of the Fonds de la Recherche Scientifique - FNRS. 

\section{Supplemental Material}
Figures S1-S6 show isomer geometries. 
Figure S7 gives the diameters of the nanoparticles optimized for each potential and for each initial starting crystal structure. 
Figure S8 shows the compute structure factors.
The optimized geometries for the clusters and the nanoparticles are available in the file \textit{Particles\_xyz.zip}.
The gab files can be opened with the software Gabedit~\cite{Allouche2011} available on sourceforge: \href{http://gabedit.sourceforge.net/}{http://gabedit.sourceforge.net/}.


%

\end{document}